# Fabrication and Characterization of PbIn-Au-PbIn Superconducting Junctions

**Nam-Hee Kim, Bum-Kyu Kim, Hong-Seok Kim and Yong-Joo Doh**[*]

*Department of Physics and Photon Science, Gwangju Institute of Science and Technology, Gwangju 61005, Korea*

**Abstract**

We report on the fabrication and measurement results of the electrical transport properties of superconductor-normal metal-superconductor (SNS) weak links, made of PbIn superconductor and Au metal. The maximum supercurrent reaches up to ~ 6 $\mu$A at $T$ = 2.3 K and the supercurrent persists even at higher temperature of $T$ = 4.7 K. Magnetic field dependence of the critical current is consistent with a theoretical fit using the narrow junction model. The superconducting quantum interference device (SQUID) was also fabricated using two PbIn-Au-PbIn junctions connected in parallel. Under perpendicular magnetic field, we clearly observed periodic oscillations of $dV/dI$ with a period of magnetic flux quantum threading into the supercurrent loop of the SQUID. Our fabrication methods would provide an easy and simple way to explore the superconducting proximity effects without ultra-low-temperature cryostats.

## 1. INTRODUCTION

Superconducting weak link, which consists of normal metal and superconducting electrodes, provides a good platform to study superconducting proximity effects.[1] Recent advancement of fabrication methods for nano-scale devices enables us to explore various quantum electronic transport phenomena in nano-hybrid mesoscopic systems combined with superconductivity.[2-8] Usually aluminum (Al) is used as a superconducting electrode because of its very long superconducting coherence length. It is, however, required to maintain the sample in a ultra-low temperature ($\ll$ 1.2 K) using a dilution refrigerator, which is caused by very low superconducting transition temperature, $T_c$, of Al ($T_c$ ~ 1.2 K). Therefore other superconducting materials with higher $T_c$ above the liquid helium temperature are necessary for a broad and relatively easy applications of the superconducting junctions.

Herein, we report the fabrications and measurements of the SNS-type Josephson junctions, which consist of superconducting PbIn electrodes and normal-metal Au electrode. The supercurrent through the PbIn-Au-PbIn junction was observed up to $T$ = 4.7 K and the maximum supercurrent increased monotonously with decreasing temperature. The superconducting quantum interference device (SQUID) was also fabricated using PbIn-Au-PbIn junctions and it exhibited a clear modulation of dynamic resistance, $dV/dI$, as a function of external magnetic field to reflect the $I_c$ modulation behavior due to superconducting phase interference effects. Our experimental study could be utilized for the applications on various superconducting devices such as a superconducting interferometer and a nano-hybrid SQUID working at the liquid-helium temperature.

## 2. EXPERIMENTALS

The PbIn-Au-PbIn weak links were fabricated in a two-step process, using conventional electron-beam lithography. Firstly, the normal-metal electrode made of Cr/Au (5/50 nm) bilayer was fabricated on the surface of Si/SiO$_2$ substrate. After that, 300 nm-thick PbIn electrodes are formed on top of the Au electrode using additional electron-beam lithography. The surface of Cr/Au electrode was treated in an oxygen-plasma chamber to remove any residual electron-beam resist before the deposition of PbIn film. The PbIn alloy film was deposited by evaporating a mixture of lead (Pb) and indium (In) pellets with a weight ratio of 1:1. Finally, we deposited 20 nm-thick Au capping layer on top of the PbIn film to avoid a probable formation of PbIn-oxide layer on it.

The superconducting transition temperature of the PbIn alloy film is obtained to be $T_c$ = 6.8 K (see Fig. 1a) and the maximum critical magnetic fields are estimated to be $H_{c,parallel}$ = 1.2 T and $H_{c,perpendicular}$ = 0.74 T (see Fig. 1b). [9] Figure 1c shows a scanning electron microscopy (SEM) image of PbIn-Au-PbIn Josephson junctions after the completion of the device fabrication process. The width of the Au electrode and the gap distance between two superconducting electrodes are found to be $w$ = 500 nm and $L$ = 220 nm, respectively. The measurement configuration is also depicted in Fig. 1c, where bias current is applied from I+ to I- and voltage difference is measured between V+ and V-.

## 3. RESULTS AND DISCUSSION

Figure 1d displays the current-voltage (*I-V*) characteristic curve obtained from sample D1 at the base temperature, $T$ = 2.3 K. It reveals a supercurrent branch without hysteresis, meaning that a superconductor-normal-superconductor (SNS) junction was well formed in the absence of a capacitive insulating interface between PbIn and Au electrodes. In our experiment, the maximum critical current, so-called a critical current ($I_c$), was obtained to be ~ 6.0 μA at the base temperature and the normal-state resistance was $R_N$ = 0.42 Ω. The inset displays the *I-V* curve measured from sample D2 at $T$ = 2.3 K, indicating $I_c$ ~ 700 nA and $R_N$ = 0.83 Ω.

Temperature-dependent $I_c$ is displayed in Fig. 2a. $I_c$ was determined from the *I-V* curve at each temperature using a linear fit crossing the zero-voltage axis. At higher temperature, $I_c$ decreases exponentially. The $I_c$ vs. *T* data are fitted well to a formula[10] of

$$I_C(T) \propto \sqrt{T} e^{-\frac{2\pi k_B T}{E_{Th}}} \qquad (eq.1)$$

where $k_B$ is the Boltzmann constant and $E_{Th}$ is the Thouless energy. We obtain $E_{Th}$ = 400 μeV as a result of fitting. Since the Thouless energy is given by $E_{Th} = \hbar D/L^2$, where $\hbar$ is the reduced Planck constant and $D$ is a diffusion constant of Au layer, we can estimate an effective gap distance to be $L^* = (\hbar D/E_{Th})^{1/2}$ = 200 nm, which is quite similar to a geometric distance $L$ = 220 nm. Here we used the Fermi velocity $v_F$ = 1.4 × 10$^8$ cm/s[11] and the mean free path $l_{mfp}$ = 50 nm to obtain $D = l_{mfp} \cdot v_F/3$ = 230 cm$^2$/s. We also note that $l_{mfp}$ was obtained from a residual resistivity of Au at $T$ = 2.3 K, which is $\rho_{Au}$ = 1.7 × 10$^{-6}$ Ωcm[12]. In addition, the superconducting coherence length is given by $\xi$ = $(\hbar D/\Delta_{PbIn})^{1/2}$ = 140 nm with the superconducting gap energy of PbIn, $\Delta_{PbIn}$ = 0.8 meV[8]. Thus our PbIn-Au-PbIn junction is in a long diffusive junction limit with satisfying $l_{\text{mfp}} \ll \xi < L$.[13]

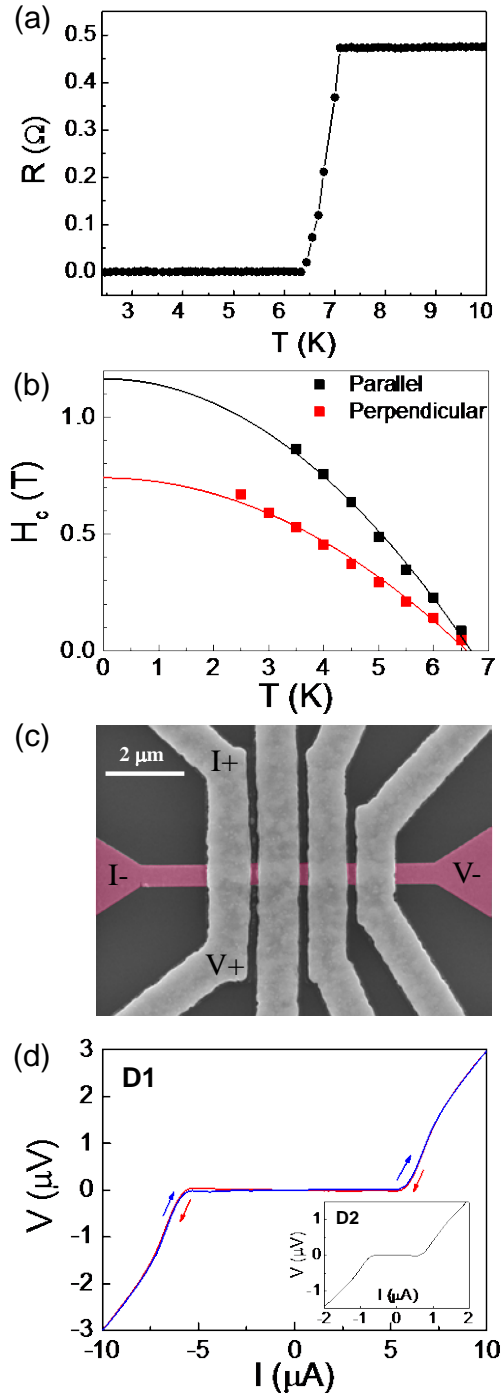

Fig. 1. (a) Temperature dependence of resistance of PbIn film. (b) Temperature-dependent critical magnetic field of PbIn film applied with parallel (black) and perpendicular (red) magnetic fields. (c) SEM image of SNS junctions with normal Au electrode (red) and superconducting PbIn electrodes (gray). (d) Current-voltage (*I-V*) curves of sample D1 at $T = 2.3$ K. Inset: *I-V* characteristic curve from sample D2 at the same temperature.

Nam-Hee Kim, Bum-Kyu Kim, Hong-Seok Kim and Yong-Joo Doh

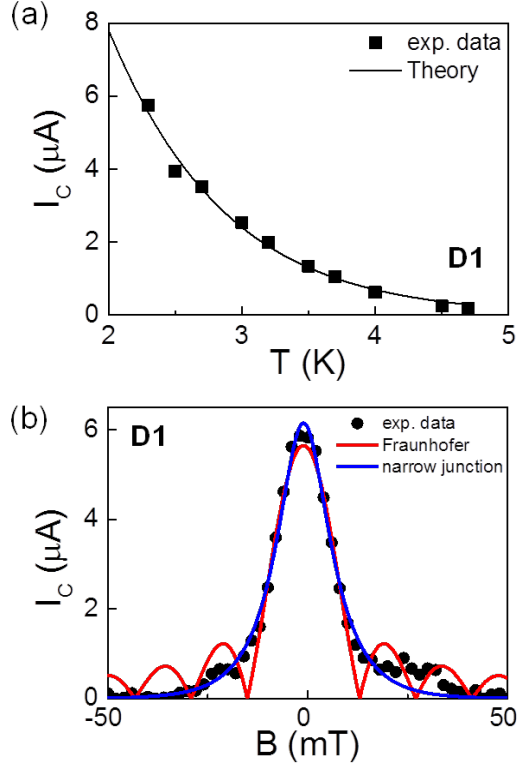

Fig. 2. (a) Temperature dependence of $I_c$. The solid line is a theoretical fit (see text). (b) Magnetic field dependence of $I_c$ at $T = 2.3$ K. The red solid line represents the ideal Fraunhofer pattern, while the blue line is a fitting results using eq.(2).

TABLE

DEVICE CHARACTERISTICS OF THE SNS JUNCTIONS AND SQUID

| Sample | $L$ (nm) | $w$ (nm) | $I_c$ ($\mu A$) at $T = 2.3$ K | $J_c$ (A/cm$^2$) |
|---|---|---|---|---|
| D1 | 220 | 540 | 6.0 | $2.2 \times 10^4$ |
| D2 | 270 | 460 | 0.70 | $3.1 \times 10^3$ |
| SQUID | 160 | 570 | 0.34 | $6.0 \times 10^2$ |

Figure 2b is the magnetic field dependence of $I_c$ at $T = 2.3$ K. It is well known that $I_c$ exhibits a periodic modulation with the application of the perpendicular magnetic field, *i.e.* Fraunhofer pattern, where the $I_c$ minima occur at the integer magnetic flux quanta through the lateral area of the junction. The theoretical prediction[14] is given by $I_c(B) = I_c(0)|\sin[\pi\Phi/\Phi_0]/(\Phi/\Phi_0)|$, where $\Phi_0$ is the magnetic flux quantum, $\Phi_0 = h/2e = 2.07 \times 10^{-15}$ Wb, and $\Phi$ is the magnetic field flux inside the junction area. However, $I_c(B)$ data (symbols) in Fig. 2b exhibit a monotonically decreasing behavior instead of a periodic modulation. The monotonous $I_c(B)$ data can be explained by the narrow junction model[15], where the junction width $w$ is comparable to or smaller than the magnetic length $\xi_H = (\Phi_0/H)^{1/2}$ and the applied $B$ field acts as a pair-breaker to suppress $I_c$.

In a narrow junction ($w < \xi_H$), the magnetic field dependence of $I_c$ is given by[16],[13]

$$eRI_C = \frac{4\pi k_B T}{r} \sum_{n=0}^{\infty} \frac{\Delta^2/(\Delta^2+\omega_n^2)}{\sqrt{2\left(\frac{\omega_n+\Gamma_H}{\varepsilon_T}\right)} \sinh\sqrt{\left(2\left(\frac{\omega_n+\Gamma_H}{\varepsilon_T}\right)\right)}} \quad \text{(eq.2)}$$

where $r = R_B/R_N$ with the barrier resistance $R_B$, $R = R_N(1+2r)$, $\omega_n = (2n+1)\pi k_B T$ is the $n$-th Matsubara energy, and $\Gamma_H =$ $De^2H^2w^2/6\hbar$ is the magnetic depairing energy. The blue solid line in Fig. 2b is the result of fitting with $r = 0.34$ and $L^* = 650$ nm, where $L^*$ is an effective junction length including the magnetic penetration effect. Here we obtained the magnetic length $\xi_H = 590$ nm, resulting in $w/\xi_H = 0.92$. Similar $I_c(B)$ behavior has been observed in other narrow junction.[17]

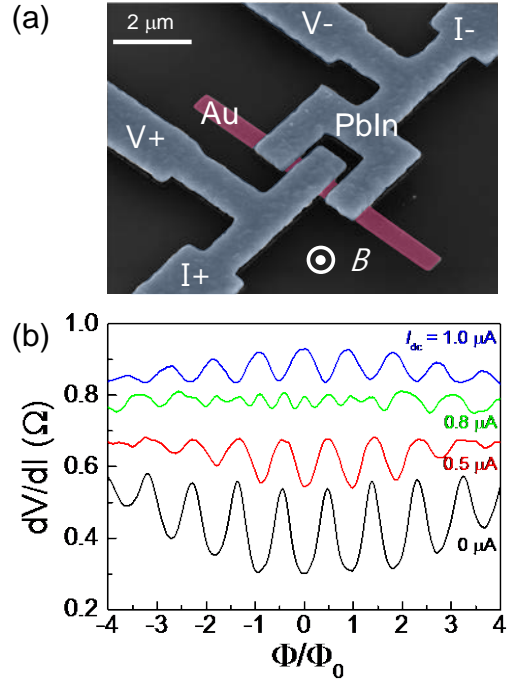

Fig. 3. (a) False-colored SEM image of SNS SQUID with Au (red) and PbIn (blue) electrodes. (b) $dV/dI$ vs. $\Phi$ curves with varying dc current $I_{dc}$. The lock-in bias current was fixed at $I_{ac} = 300$ nA.

Two SNS junctions connected in parallel can form a SQUID, as depicted in Fig. 3a with the measurement configuration. Under the application of perpendicular magnetic field $B$, periodic modulation of dynamic resistance ($dV/dI$) was observed using lock-in measurement combined with dc ($I_{dc}$) and ac ($I_{ac}$) bias currents. The oscillation periodicity is obtained to be $H^* = 6.0$ Oe, resulting in $\Phi^* = 8.6 \times 10^{-16}$ Wb = $0.42\ \Phi_0$ using a geometric inner-loop area of $A = 1.4\ \mu m^2$. The difference of $\Phi^*$ from an expected value of $\Phi_0$ is attributed to an effective loop area considering the London penetration depth $\lambda_L$. We can estimate $\lambda_L$ from an effective loop area, $A^* = \Phi_0/H^* = 3.4\ \mu m^2$, resulting in $\lambda_L = 470$ nm, which is close to the value obtained from other nanowire-based SQUID.[8]

In conclusion, we have established an easy method for fabricating an SNS weak link and applied it to fabricate SQUID. We observed a dissipationless supercurrent flowing through the PbIn-Au-PbIn junction above the liquid-helium temperature. The $I_c$ modulation with an external magnetic field resembles the one expected from the narrow junction model rather than the Fraunhofer pattern. The periodic $dV/dI$ oscillations are clearly observed in a SQUID consisting of two PbIn-Au-PbIn junctions connected in parallel. Our experimental results would provide a useful and convenient way for exploring the superconducting phase interference effects near the liquid-helium temperatures.

## ACKNOWLEDGMENT

This work was supported by the National Research Foundation of Korea through the Basic Science Research Program (Grant No. 2015R1A2A2A01006833).

# REFERENCES

1. Likharev, K.K., *Superconducting weak links.* Reviews of Modern Physics, 1979. **51**(1): p. 101-159.
2. Doh, Y.-J., et al., *Tunable supercurrent through semiconductor nanowires.* Science, 2005. **309**(5732): p. 272-275.
3. Doh, Y.-J., et al., *Andreev Reflection versus Coulomb Blockade in Hybrid Semiconductor Nanowire Devices.* Nano Letters, 2008. **8**(12): p. 4098-4102.
4. Jung, M., et al., *Superconducting junction of a single-crystalline Au nanowire for an ideal Josephson device.* ACS Nano, 2011. **5**(3): p. 2271-2276.
5. Lee, G.-H., et al., *Electrically tunable macroscopic quantum tunneling in a graphene-based Josephson junction.* Physical Review Letters, 2011. **107**(14): p. 146605.
6. Jeong, D., et al., *Observation of supercurrent in PbIn-graphene-PbIn Josephson junction.* Physical Review B, **2011**. **83**(9): p. 094503.
7. Choi, J.-H., et al., *Complete gate control of supercurrent in graphene p–n junctions.* Nature Communications, **2013**. **4**: p. 2525.
8. Kim, H.-S., et al., *Gate-tunable superconducting quantum interference devices of PbS nanowires.* Appl. Phys. Express, **2016**. **9**: p. 023102.
9. Park, S.-I., et al., *Characterizing Pb-based superconducting thin films.* Prog. Supercond. Cryogenics , 2014. **16**(4): p. 36-39.
10. Kresin, V.Z., *Josephson Current in Low-Dimensional Proximity Systems and the Field-Effect.* Physical Review B, 1986. **34**(11): p. 7587-7595.
11. Lee, G.-H., et al., *Ultimately short ballistic vertical graphene Josephson junctions.* Nature Communications, 2015. **6**: p. 6181.
12. Ashcroft, N.W. and N.D. Mermin, *Solid state physics.* 1976, USA: Brooks/Cole.
13. Bergeret, F.S. and J.C. Cuevas, *The vortex state and Josephson critical current of a diffusive SNS junction.* Journal of Low Temperature Physics, 2008. **153**(5-6): p. 304-324.
14. Tinkham, M., *Introduction to Superconductivity.* 1996: Dover Publications.
15. Cuevas, J.C. and F.S. Bergeret, *Magnetic interference patterns and vortices in diffusive SNS junctions.* Physical Review Letters, 2007. **99**(21): p. 217002.
16. Hammer, J.C., et al., *Density of states and supercurrent in diffusive SNS junctions: Roles of nonideal interfaces and spin-flip scattering.* Physical Review B, 2007. **76**(6): p. 064514.
17. Kim, B.-K., et al., *Very Strong Superconducting Proximity Effects in PbS Semiconductor Nanowires.* ArXiv : 1607.07151, 2016.